\documentstyle[12pt]{article}

\newcommand{\eq}{\begin{equation}}
\newcommand{\en}{\end{equation}}
\newcommand{\bea}{\begin{eqnarray}}
\newcommand{\eea}{\end{eqnarray}}
\newcommand{\spz}{\hspace{0.7cm}}

\newcommand{\virg}{\spz,\spz}

\newcommand{\wave}[1]{\mbox{\raisebox{-.6ex}{$\stackrel{\displaystyle{\sim}}
                     {\scriptstyle{x \rightarrow {#1}}}$}}}

\newcommand{\D}{\Delta}
\newcommand{\SZ}{{\bf SZ}}

\newcommand{\ih}{\hat{\imath}}
\newcommand{\jh}{\hat{\jmath}}
\newcommand{\kh}{\hat{k}}
\newcommand{\uno}{\hat{1}}
\newcommand{\due}{\hat{2}}
\newcommand{\corr}[4]{\langle {#1}~{#2}~{#3}~{#4} \rangle}
\newcommand{\half}{\frac{1}{2}}

\newcommand{\NP}[1]{Nucl.\ Phys.\ {\bf #1}}
\newcommand{\PL}[1]{Phys.\ Lett.\ {\bf #1}}

\newcommand{\CMP}[1]{Comm.\ Math.\ Phys.\ {\bf #1}}

\newcommand{\IJMP}[1]{Int.\ J.\ Mod.\ Phys.\ {\bf #1}}
\newcommand{\JETP}[1]{Sov.\ Phys.\ JETP {\bf #1}}
\newcommand{\TMP}[1]{Teor.\ Math.\ Phys.\ {\bf #1}}

\begin{document}

\renewcommand{\thefootnote}{\fnsymbol{footnote}}

\newpage
\setcounter{page}{0}

\vskip 1cm
\begin{center}
{\bf ON THE POSSIBILITY OF $Z_N$ EXOTIC SUPERSYMMETRY IN TWO
     DIMENSIONAL CONFORMAL FIELD THEORY}\\
\vskip 1.8cm
{\large F.\ Ravanini}\\
\vskip .7cm
{\em Service de Physique Th{\'e}orique, C.E.A. - Saclay \footnote{
     Laboratoire de la Direction des Sciences de la Mati{\`e}re du
     Commissariat \`a l'Energie Atomique} \\
     Orme des Merisiers, F-91190 Gif-sur-Yvette, France\\
     and\\
     I.N.F.N. - Sez. di Bologna, Italy}
\end{center}
\vskip 1cm

\renewcommand{\thefootnote}{\arabic{footnote}}
\setcounter{footnote}{0}

\begin{abstract}
We investigate the possibility to construct extended parafermionic conformal
algebras whose generating current has spin $1+\frac{1}{K}$, generalizing the
superconformal (spin 3/2) and the Fateev Zamolodchikov (spin 4/3) algebras.
Models invariant under such algebras would possess $Z_K$ exotic supersymmetries
satisfying (supercharge)$^K$ = (momentum). However, we show that for $K=4$
this new algebra allows only for models at $c=1$, for $K=5$ it is a
trivial rephrasing of the ordinary $Z_5$ parafermionic model, for $K=6,7$
(and, requiring unitarity, for all larger $K$) such algebras do not exist.
Implications of this result for existence of exotic supersymmetry in two
dimensional field theory are discussed.
\end{abstract}
\vskip .3cm
\begin{flushright}
Saclay preprint SPhT/91-121\\
August 1991
\end{flushright}
\vskip .3cm
Submitted for publication to Int.J.Mod.Phys. A
\newpage

\section{Introduction}
Conformal Field Theory (CFT) in two dimensions (2D)~\cite{bpz,reviews}
can give information on
the general structure of the space of {\it all} 2D Quantum Field Theories
(QFT). Indeed, each reasonable QFT must possess ultraviolet (UV) as
well as infrared (IR) fixed points for which the Callan-Symanzik $\beta$
function
is zero, thus showing scale invariance. Following Polyakov~\cite{poly} it is
conceivable that all scale invariant QFT also possess conformal invariance.
Hence the UV and IR limits of any 2D QFT must be described by suitable CFT's.

Moreover, if
a QFT has some particular symmetry preserved all along the renormalization
flow, this symmetry should exhibit itself at the UV and IR points too. For
example consider an $N=1$ supersymmetric theory. Its action is invariant under
transformations by a spin 1/2 charge $Q$ (the so called {\em supercharge})
such that $Q^2=P$, where $P$ is the
(conserved) total momentum. Corresponding to this charge $Q$ there is a
conserved
current of spin 3/2 that in the UV limit becomes the well known $G(z)$ current
enlarging the conformal symmetry to an $N=1$
superconformal one. Thus the UV limit of such a theory must be a
superconformal model, minimal or not.
In a certain sense, we could say that existence of a superconformal algebra
guarantees
the existence of reasonable UV limits for 2D $N=1$ supersymmetric QFT's.

One can ask if such a structure can be generalized to models having, say,
$Z_3$ graded supersymmetry, i.e. models whose action is invariant under
transformations $Q$, $Q^{\dagger}$ such that $Q^3=Q^{\dagger~3}=P$. If so,
the UV limit of these $Z_3$
supersymmetric models is described by appropriate ``$Z_3$ exotic''
superconformal
models, invariant under an algebra that generalizes the superconformal one
to the case of $Z_3$ gradation. As $P$ has dimension 1, $Q$ and $Q^{\dagger}$
must have
dimension 1/3, and the corresponding conserved current must be of dimension
4/3. Such an algebra extending the Virasoro algebra by means of a couple of
conjugated currents $A(z)$, $A^{\dagger}(z)$ of spin (and dimension) 4/3 has
been investigated by Fateev and Zamolodchikov~\cite{fz2}.

These $Z_2$ (Susy) and $Z_3$ (spin 4/3) algebras, that we shall call in the
following $\SZ_2$ and $\SZ_3$ respectively, show some well known common
structure. First of all, both allow for a series of unitary minimal models,
accumulating to $c=3/2$ and to $c=2$ respectively~\cite{fz2,fqs2}.
For $c$ larger than these
values there is still a continuum of non-minimal models. The set of all $\SZ_2
\otimes \overline{\SZ}_2$
invariant models, minimal or not, is the set of all UV or IR fixed point of
2D QFT invariant under $N=1$ supersymmetry; the same set for $\SZ_3$ represents
all fixed points of QFT invariant under $Z_3$ exotic supersymmetry.
Each minimal model in the two series can be
perturbed by some relevant scalar operator contained in its Kac table.
The {\em least relevant operator} (that with
conformal dimension closer to 1) leads to two different behaviours: for
negative values of the perturbing coupling constant the model is massive and
integrable, and its scattering matrix is known~\cite{abl}; for
positive coupling constant the perturbation defines a massless flow that has
been shown, at least by perturbative arguments, to have a non trivial IR limit
also belonging to
the same series~\cite{kms,sotkov}. The $Z_2$ or $Z_3$ supersymmetry
is preserved
along the flow, i.e. one can define non-local charges $Q$ in the perturbed
model, such that $Q^2=P$ or $Q^3=P$ respectively.

In particular, when one perturbs by the least relevant operator the model
in the minimal series having the lowest central charge $c$, one can show that
still there is a conserved current $Q$ in the perturbed model with $Q^2=P$ or
$Q^3=P$. For negative coupling constant the scattering matrix of the massive
model is known~\cite{zam} and is indeed invariant under such a charge $Q$. For
positive coupling the models are know to flow to usual $Z_2$ (Ising) and
$Z_3$ (Potts) models respectively. As these IR limits do not have $Q$
invariance any more, one concludes~\cite{kms,zam} that along the flow there is
spontaneous supersymmetry breaking. In the $Z_2$ case, it has been
shown~\cite{kms} that
the resulting goldstino is a field that in the IR limit evolves in the
spin 1/2 fermion of the Ising model. Similarly, in the $Z_3$ case,
Zamolodchikov~\cite{zam} has argued that the goldstini fields
corresponding to the
broken $A(z)$, $A^{\dagger}(z)$ currents should become, in the IR limit, the
couple of
3-state Potts parafermions of spin 2/3. Notice that in both cases
the spin of the broken current and that of the resulting goldstino sum up to
2 (any better understanding of this fact should be welcome).
It has also been observed that in both cases the scattering matrix of the
negative coupling massive model coincides with the Boltzmann weights of the
corresponding $Z_2$ Ising or $Z_3$ Potts models~\cite{zam}.

In this paper we address the problem to generalize
these examples to $Z_K$ graded exotic superconformal algebras, $\SZ_K$
for short. If such
algebras exist, they can be the base for the construction of $\SZ_K \otimes
\overline{\SZ}_K$ invariant models. Then one can address the problem of
perturbing these models by their least relevant operator thus getting examples
of massive and massless non-conformal $Z_K$ exotic supersymmetric models.
In particular, the picture valid for $Z_2$ and $Z_3$ should naively suggest
that suitably perturbing the $\SZ_K$ model with lowest central charge, one
could get a case of spontaneously broken $Z_K$ exotic supersymmetry, whose
goldstino, in the IR limit, could describe the usual $Z_K$ parafermion of the
$Z_K$ Ising model. Boltzmann weights for the $Z_K$ Ising models are
known~\cite{fzboltz}. Assuming that they can be as well used as scattering
matrices of some 2D QFT, Bernard and Pasquier~\cite{bp} have shown that
they are indeed
invariant under transormations $Q$ such that $Q^K=P$. Thus, they are the
natural candidates of an eventual massive model obtained by suitably deforming
the lowest $c$ $\SZ_K$ model.

If conversely the $\SZ_K$ models do not exist, namely because the $\SZ_K$
algebras are inconsistent for some $K$, then no UV limit
can be defined for a $Z_K$ exotic supersymmetric theory, and, as a reasonable
QFT must have an UV limit, we can conclude that no $Z_K$ exotic supersymmetry
exists for that $K$ at all, and the Boltzmann weights of~\cite{bp} cannot be
used as scattering matrices of a QFT.

The interest of searching possible $\SZ_K$ algebras is even more general: in
appendix A of~\cite{fz}, Fateev and Zamolodchikov describe the most general
$Z_K$ symmetric parafermionic algebra, of which the usual $Z_K$ parafermions
are a particular case. All $Z_K$-invariant conformal models should have some
realization of this general $Z_K$ algebra somewhere hidden in their operator
product expansion (OPE) algebra. Thus, knowledge of possible associative
$Z_K$ symmetric algebras (and of their representations) should help in the
classification of all conformal models having $Z_K$ symmetry. $\SZ_K$ algebras
explored in the present paper are one among the many possibilities described
in~\cite{fz}.

\section{The $Z_K$-superconformal algebras}

We begin our investigation by giving the general form of the $\SZ_K$ algebras
we are interested in. We proceed by direct generalization of the known $Z_2$
and $Z_3$ cases.
The $\SZ_2$ algebra has the simple $Z_2$-graded fusion rules $\psi \psi = 1$,
and the $\SZ_3$ one is described by $Z_3$-graded fusion rules $\psi \psi =
\psi^{\dagger}$, $\psi \psi^{\dagger}=1$. Requirement of $Z_K$-gradation
of the $\SZ_K$ algebra means considering a set of $Z_K$ symmetric fusion rules
for the currents:
\eq
\psi_i \psi_j = \psi_{i+j}
\spz,\spz (\psi_0=1,~~\psi_i^{\dagger}=\psi_{K-i})
\label{fusion}
\en
Here and in the following $i,j,k,...$ indices are always to be taken modulo
$K$ and we introduce the notation $\ih=K-i$.
We are interested in $N=1$ supersymmetry, i.e. we require that
for a given conformal dimension $\Delta_k$
there can be only {\em one} couple of currents $\psi_k$ and $\psi_k^{\dagger}$.
Furthermore, we require that
no currents of spin one are present as secondaries in the family of the
identity, otherwise they would form the Kac-Moody algebra of a
continuous internal symmetry, while we are interested in the case
where no symmetry additional to the $Z_K$-susy is postulated. We also require
that the only current of dimension two appearing in the identity family is the
stress-energy tensor. Currents of higher spin 3,4,5... are allowed, which
means that the identity family is not necessarily that of pure Virasoro
algebra, it can as well contain currents generating some W-algebra. All these
requirements fix the form of the operator product expansion (OPE) algebra to
be the following ($z_{12}=z_1-z_2$ and $\jh\not= i$
\eq
\begin{array}{l}
\psi_i(z_1)\psi_j(z_2)=C_{ij}z_{12}^{\alpha_{ij}} \left[
                     \psi_{i+j}(z_2) + z_{12}\frac{\alpha_{ij}+2\D_i}
                     {2\D_{i+j}}
                     \partial_{z_2}\psi_{i+j}(z_2) +O(z_{12}^2) \right]\\
\psi_i(z_1)\psi_i^{\dagger}(z_2)= z_{12}^{-2\D_i}
                         \left[ 1+z_{12}^2\displaystyle{\frac{2\D_i}{c}}T(z_2)
                         +O(z_{12}^3) \right] \\
T(z_1)\psi_i(z_2)=\displaystyle{\frac{\D_i\psi_i(z_2)}{z_{12}^2}
                  +\frac{\partial_{z_2}\psi_i(z_2)}{z_{12}}}
                    +O(1) \\
T(z_1)T(z_2)=\displaystyle{\frac{c/2}{z_{12}^4}+
             \frac{2T(z_2)}{z_{12}^2}+\frac{\partial_{z_2}T(z_2)}{z_{12}}}
             +O(1)
\end{array}
\label{ope}
\en
where we introduced the useful notation $\alpha_{ij}=\D_{i+j}-\D_i-\D_j$.
The last two equations show that Virasoro algebra with central charge $c$ is a
subalgebra of $\SZ_K$ algebra and that $\psi_i$'s are Virasoro
primary fields of left conformal dimension $\D_i$. As they are to be
conserved currents, their right conformal dimension $\bar{\D}_i$ must be zero.
Hence the spin of the current $\psi_i$, as well as its full conformal
dimension, is given by $\D_i$. Of course there will be a ``right'' algebra
of currents $\bar{\psi_i}(\bar{z})$ and $\bar{T}(\bar{z})$ pertaining the
right chiral part of the models and commuting with the \{$\psi_i,T$\} algebra,
so that the models will be invariant under $\SZ_K \otimes \overline{\SZ}_K$
symmetry.
All the considerations in the following will be done for the left algebra and
apply as well to the right one.

The spins $\D_k$ can not take arbitrary values. As explained in Appendix A
of~\cite{fz}, or equivalently using the techniques of~\cite{christe,vafa}, it
is possible to show that the most general value for $\D_k$ compatible with
the fusion rules (\ref{fusion}) is given by
\eq
\D_k = \frac{pk(K-k)}{K} + M_k
\en
where $M_k\in{\bf Z}$, $M_{\kh}=M_k$, $M_0=0$ and $p$ can be integer if $K$ is
odd and integer or half-integer if $K$ is even.
Furthermore, in the case of $\SZ_K$ algebra, we must require that one of the
currents
$\psi_k$, say $\psi_1$, has spin $\D_1=1+1/K$, in order to be able to define
a conserved charge $Q$ such that $Q^K=P$. This fixes $M_1=M_{\uno}=2$ and
$p=-1$, hence the formula for $\D_k$ we shall assume in the following is
\eq
\D_k = M_k - \frac{k(K-k)}{K} \spz,\spz M_0=0 \spz M_1=2 \spz M_{\kh}=M_k
\label{delta}
\en
where the integers $M_k$ have to be constrained by the allowed
behaviours of correlation functions near their singularities (see below).

The structure constants $C_{ij}$ are chosen such that all non-zero two point
functions are normalized to 1 ($C_{i\ih}=1$), and enjoy full symmetry,
i.e. defining
\eq
C_{ij}=Q_{i,j}^{i+j}=Q_{i,j,N-i-j}
\en
full symmetry of the symbols $Q_{i,j,k}$ must be required. This restricts the
number of indipendent structure constants. Moreover charge conjugation
symmetry implies $C_{ij}=C_{\ih\jh}^*$.

\section{Associativity}

The most important requirement on $\SZ_K$ algebras is their associativity or,
equivalently, duality of the 4-point
correlation functions of fields $\psi_i$.
As fields $\psi_i(z)$ do not depend on $\bar{z}$,
their 4-point functions will have dependence on $z_1,...,z_4$ only, and not
on $\bar{z}_1,...,\bar{z}_4$.
Invariance under the projective group SL(2,{\bf C}) implies that one can
choose 3 of the 4 positions in any 4-point function as 0,1 and $\infty$
so that it depends essentially only on one
projective invariant variable $x$, the so called {\em anharmonic
ratio}
\eq
\langle 0|\psi_l (\infty) \psi_k (1) \psi_i (x) \psi_j(0) |0\rangle
= G_{ij}^{kl}(x)
\en
$Z_K$ invariance forces this correlation function to be 0 if $i+j+k+l \not= 0
\bmod K$.

The requirement of duality on the 4-point function can be put
as conditions on $G_{ij}^{kl}(x)$
\eq
G_{ij}^{kl}(x) = G_{ik}^{jl}(1-x) = e^{\pi i \alpha_{ik}}
x^{-2\Delta_i} G_{il}^{kj}(1/x)
\label{duality}
\en
The phase in the last equality comes from the braiding of the (semilocal)
fields $\psi_i$ and
$\psi_k$ necessary to bring $x$ close to $\infty$. Due to the mutual
semilocality of fields~\cite{fz},
$G_{ij}^{kl}$ is not in general a single valued function. Its expansion in
blocks must reproduce its monodromy properties, which in turn can be read from
the OPE's (\ref{ope}). Obviously, blocks for the right chiral
part of the
correlation function are trivially equal to 1. Moreover, in each channel
there is only {\em one} possible exchanged family and
therefore only one block:
\eq
\begin{array}{lll}
G_{ij}^{kl}(x)   &=& C_{ij} C_{kl} {\cal F}_{ij}^{kl}(x)   \\
G_{ik}^{jl}(1-x) &=& C_{ik} C_{jl} {\cal F}_{ik}^{jl}(1-x) \\
G_{il}^{kj}(1/x) &=& C_{il} C_{kj} {\cal F}_{il}^{kj}(1/x)
\end{array}
\en
The behaviour of the blocks at $x=0,1,\infty$ can be easily
inferred from the OPE's
\eq
\begin{array}{lll}
{\cal F}_{ij}^{kl}(x) & \wave{0} &
   x^{\alpha_{ij}} \displaystyle{\sum_{n=0}^{\infty}} c_n x^n\\
{\cal F}_{ik}^{jl}(1-x) & \wave{1} &
   (1-x)^{\alpha_{ik}} \displaystyle{\sum_{n=0}^{\infty}} d_n (1-x)^n\\
{\cal F}_{il}^{kj}(1/x) & \wave{\infty} &
   \left( \frac{1}{x} \right)^{\alpha_{il}} \displaystyle{\sum_{n=0}^{\infty}}
   h_n x^{-n}
\end{array}
\label{beh}
\en
The series expansion are convergent in a neighborhood of $0,1,\infty$
respectively. Some of
the coefficients $c_n,d_n,h_n$ can be computed from the information contained
in the OPE's (\ref{ope}).
In particular $c_0,d_0,h_0$ are always guaranteed to be equal
to 1. Moreover, if $i\not= \jh$
\eq
{\cal F}_{ij}^{kl}(x)=x^{\alpha_{ij}}\left( 1+\frac{(\alpha_{ij}+2\D_i)
(\alpha_{ij}+2\D_j)}{2\D_{i+j}}x + O(x^2)\right)
\label{block2}
\en
while if $i=\jh$
\eq
{\cal F}_{i\ih}^{k\kh}(x)=x^{-2\D_i}\left( 1+\frac{2\D_i\D_k}{c}x^2+O(x^3)
\right)
\label{block}
\en
The behaviour for the blocks in eq.(\ref{beh}) reproduces the correct monodromy
properties of the multivalued correlation function.
The blocks are (up to the branch singularity in the leading
factor) locally holomorphic functions of $x$. They can be analytically
continued in the whole plane, excluding the branch points.
Closure under analytic continuation requires~\cite{christe,christe-thesis}
\eq
\alpha_{ij}+\alpha_{ik}+\alpha_{il}-2\D_i=-R
\label{R}
\en
where $R$ is a non negative integer. This in turn fixes the general form of
blocks to be
\eq
\begin{array}{lll}
{\cal F}_{ij}^{kl}(x)   &=& x^{\alpha_{ij}}(1-x)^{\alpha_{ik}}P(x) \\
{\cal F}_{ik}^{jl}(1-x) &=& (1-x)^{\alpha_{ik}}x^{\alpha_{ij}}Q(1-x)  \\
{\cal F}_{il}^{kj}(1/x) &=& x^{-\alpha_{il}}(1-x^{-1})^{\alpha_{ik}}T(1/x) \\
                        &=& e^{-\pi i \alpha_{ik}}x^{2\D_i}x^{\alpha_{ij}}
                            (1-x)^{\alpha_{ik}}x^RT(1/x)
\end{array}
\en
where $P(x)=\sum_{n=0}^RP_nx^n$, $Q(x)=\sum_{n=0}^RQ_nx^n$ and $T(x)=
\sum_{n=0}^RT_nx^n$ are polynomials of degree $R$ in $x$.
In the last equality eq.(\ref{R}) has been used.
To avoid heavy notation we dropped indices $i,j,k,l$ from the polynomials,
but it must be bared in mind that they, as well as the integer $R$,
are specific of the particular
correlation function $G_{ij}^{kl}(x)$. Normalization of the blocks implies
$P_0=Q_0=T_0=1$.

The duality requirement (\ref{duality}) can then be simply translated in
conditions on these polynomials
\eq
\begin{array}{lll}
\mbox{{\em st}-duality:}&~\Rightarrow~&
                        C_{ij}C_{kl}^*P(x)=C_{ik}C_{jl}^*Q(1-x) \\
\mbox{{\em su}-duality:}&~\Rightarrow~&
                        C_{ij}C_{kl}^*P(x)=C_{il}C_{kj}^*x^RT(1/x)
\end{array}
\en
These equations can take a particularly simple form when they are considered
for some special cases. Here and in the following we use the notation
$G_{ij}^{kl}=\corr{i}{j}{k}{l}$.
Remembering that $C_{i,\ih}=1$, we first consider
the correlation functions $\corr{i}{\ih}{k}{\kh}$. The constraints
\eq
\frac{2\D_i\D_k}{c}=P_2-\half(\alpha_{ik}^2+\alpha_{ik})
\label{c}
\en
and
\bea
P_1&=&\alpha_{ik} \nonumber \\
Q_1&=&\frac{(\alpha_{ik}+2\D_i)(\alpha_{ik}+2\D_k)}{2\D_{i+k}}+\alpha_{ij}\\
T_1&=&\frac{(\alpha_{i\kh}+2\D_i)(\alpha_{i\kh}+2\D_k)}{2\D_{i-k}}+\alpha_{ik}
\nonumber
\eea
obtained comparing the blocks with the expansions (\ref{block2},\ref{block})
can be conveniently used, together with the duality constraints
\eq
|C_{ik}|^2=P(1)=\sum_{n=0}^R P_n \virg
Q_n=\frac{1}{|C_{ik}|^2}\sum_{p=n}^R \left( \begin{array}{c} n\\p \end{array}
\right) P_p
\label{q}
\en
\eq
|C_{i,\kh}|^2=P_n \virg
T_n=\frac{1}{|C_{i,\kh}|^2}P_{R-n}
\label{t}
\en
to fix as much as possible the parameters. In
particular for $R=0$ it should be $\alpha_{ik}=0$ and then $c=\infty$.
Hence if at least one of the correlation functions $\corr{i}{\ih}{i}{\ih}$ has
$R=0$ the whole algebra is unconsistent.

In the case $i=k$, the correlation function $\corr{i}{\ih}{i}{\ih}$ has even
more constraints. Indeed, here {\em su}-duality
brings again to the same block, hence the polynomial $T(x)$ is equal to
$P(x)$, and we have the condition
\eq
P(x)=x^RP(1/x) ~\Rightarrow~ P_{R-n}=P_n
\en
It is easy to convince oneself that if $R\leq 3$ the polynomial is
completely fixed. For $R=1$ the only way to avoid
unconsistency is to have $\alpha_{ik}=1$.
In the $R=2,3$ cases eq.(\ref{c}) can be conveniently used to fix $c$. For
higher values of $R$ (\ref{c}) can be used to
express the coefficient $P_2$ in terms of $c$.

Another case of strongly constrained correlation function is that of four
equal fields $\corr{i}{i}{i}{i}$. In this case the blocks in all the three
channels coincide and there is only one polynomial whose coefficients are
constrained by the equations $P(x)=x^RP(1/x)$ and $P(x)=P(1-x)$. They have
no solution for $R$ odd, for $R$ even the coefficients $P_n$
are fully determined if $R<6$.

A last remark is in order about conformal
dimensions: we have seen in the previous section that they are determined
up to integers $M_k$. Now, conformal dimensions enter in the determination
of $R$, eq.(\ref{R}). The
fundamental requirement that for all 4-point functions $R$ must be a non
negative integer can then be translated into a set of inequalities to be
satisfied by the integers $M_k$, thus strongly selecting, as we shall see
in the next section, the possible choices of conformal dimensions.

The strategy to study a given algebra will then be the following:
\begin{enumerate}
\item
identify the parameters in the algebra. These are $c$ and the independent
structure constants. Moreover, spin of fields will eventually depend on
a set of integers $M_k$ as in eq.(\ref{delta}).
\item
list all the non-zero 4-point functions and compute $R$ for each of them.
Imposing $R\geq 0$ the integers $M_k$ can be selected to a few possible
choices.
\item
For each choice consider the functions $\corr{i}{\ih}{k}{\kh}$. If at least
one of them has $R=0$, discard the choice. Else, identify the $\corr{i}{\ih}
{k}{\kh}$ functions with lower values of $R$ and, after having constrained
as much as possible the coefficients of the polynomials using duality,
try to fix $c$ and/or some $C_{ij}$ through eqs.(\ref{c},\ref{q},\ref{t}).
If different
correlation functions provide unconsistent values of $c$ or $C_{ij}$,
discard the choice.
\item
Otherwise use the values so obtained to fix as much as possible the other
correlation functions and check all the possible duality constraints. If
somewhere some unconsistency appears, discard the choice. If instead the
choice passes all the checks it defines a consistent associative algebra.
If some parameter is left free, the algebra allows for an infinity of
associative realizations, one for each value of the parameter.
\item
redo steps 3 and 4 for all the choices selected by step 2.
\end{enumerate}
Next section will illustrate this procedure on some simple examples.

\section{$\SZ_K$ algebras for $K \leq 7$}

To illustrate the general theory of the previous section, let us study some
particular case in detail. We shall discover that even for low values of
$K$ some interesting surprises arise.

\vskip 0.3cm
{\bf $\SZ_2$ algebra -} In the case $K=2$, i.e. usual $N=1$ superconformal
algebra there is only one fermionic field $\psi_1$ of spin 3/2.
No non trivial structure constant appear and the only parameter in the
algebra is $c$.
There is only one non-trivial 4-point function $\corr{1}{1}{1}{1}$
with $R=6$. Crossing symmetry fixes it up
to a free parameter:
\eq
\corr{1}{1}{1}{1} = x^{-3}(1-x)^{-3}(1-3x+Px^2+\frac{9-2P}{3}x^3+Px^4-3x^5
                    +x^6)
\en
that can be re-expressed in terms of $c$ as $P=(15c-3)/2c$.
No other restriction can be found on the algebra. Therefore we conclude
that $\SZ_2$ is associative for any value of $c$. Unitarity~\cite{fqs2}
will then
restrict to $c=\frac{3}{2}\left( 1-\frac{8}{m(m+2)}\right)$ or $c\geq 3/2$.

\vskip 0.3cm
{\bf $\SZ_3$ algebra -} The case $K=3$, namely the spin 4/3 algebra of Fateev
and Zamolodchikov~\cite{fz2}, has two free parameters: $c$ and the structure
constant $C_{1,1}=\lambda$.
On the other hand, there is only one non trivial 4-point
function $\corr{1}{\uno}{1}{\uno}$, for which
$R=4$. Crossing symmetry then
determines the 4-point function up to a free parameter $P$
\eq
\corr{1}{\uno}{1}{\uno} =
x^{-8/3}(1-x)^{-4/3}(1-\frac{4}{3}x+Px^2-\frac{4}{3}x^3+x^4)
\en
Both $c$ and $\lambda$ can be computed in terms of $P$ by use of
eqs.(\ref{c},\ref{q}).
Eliminating $P$ in the result we reobtain the well known relation computed
by Fateev and Zamolodchikov $9c|\lambda|^2=4(8-c)$. Hence the $\SZ_3$ algebra
has only one free parameter, that can be chosen to be $c$. Unitarity~\cite{gs}
will fix it to $c=2\left(1-\frac{12}{m(m+4)}\right)$ or $c\geq 2$.

\vskip 0.3cm
{\bf $\SZ_4$ algebra -} The first new case $K=4$ has parafermions $\psi_1$,
$\psi_2$ and $\psi_3=\psi_1^{\dagger}$
of dimensions $\D_1=\D_3=5/4$ and $\D_2=M_2-1$ respectively. The
algebra has two parameters: $c$ and the structure constant $C_{1,1}=\lambda$.
There are four non trivial 4-point functions. The requirement that for each of
them $R\geq 0$ amounts to a set of inequalities in $M_2$:
\begin{eqnarray*}
\mbox{from } \corr{1}{\uno}{1}{\uno} &\Rightarrow& M_2 \leq 6\\
\mbox{from } \corr{2}{2}{2}{2} &\Rightarrow& M_2 \geq 1\\
\mbox{from } \corr{1}{\uno}{2}{2} &\Rightarrow& M_2 \geq 1\\
\mbox{from } \corr{1}{1}{1}{1} &\Rightarrow& M_2 \leq 2\\
\end{eqnarray*}
We see that these, together with the fact that $\D_2$ must be different from
zero to avoid doubling of the vacuum, imply $M_2=2$ and therefore $\D_2=1$.
The correlation function $\corr{2}{2}{2}{2}$ with $R=4$ is completely fixed
by crossing symmetry
\eq
\corr{2}{2}{2}{2} = x^{-2}(1-x)^{-2}(1-2x+3x^2-2x^3+x^4)
\en
and can be conveniently used to fix $c=1$ by use of eq.(\ref{c}).
The other correlation function $\corr{1}{\uno}{1}{\uno}$ has $R=4$.
Duality should fix it up to a
parameter depending on $c$. The value $c=1$ fixes this parameter, and the
correlation function reads
\eq
\corr{1}{\uno}{1}{\uno} = x^{-5/2}(1-x)^{-3/2}(1-\frac{3}{2}x+\frac{7}{2}x^2
                    -\frac{3}{2}x^3+x^4)
\en
Eq.(\ref{q}) then allows to compute the structure constant (up to an
inessential phase that we fix to 1) as $\lambda=\sqrt{5/2}$.
One can check that the
remaining correlation functions are compatible with this fixing of the
parameters.

The surprising result here is that $c$ is no more free. $\SZ_4$ does not allow
for a series of minimal models at different $c$ plus continuum,
but rather for a series of models
at $c=1$, that can be indeed easily identified with points on the
gaussian and orbifold lines
for suitable values of the compactification radius. In a certain sense,
the series of models
present in $\SZ_2$ and $\SZ_3$ cases are here ``squeezed'' to the $c=1$ lines.
The models can still ``flow'' from one another, but the perturbing operator
is now the limiting case of least relevant operators, namely the marginal
operator that allows to move along the $c=1$ lines. All the $\SZ_4$ invariant
models are connected by this marginal operator to one of the two modular
invariant solutions of $Z_4$ parafermion (the diagonal one lying on the
orbifold line). One could then still speculate about $\SZ_4$ spontaneous
symmetry breaking with goldstini given by $Z_4$ parafermions. This picture is
however somewaht delicate here:
the current $\psi_1$ of spin 5/4 seems to break nicely to give a paragoldstino
of spin 3/4 (again $5/4 + 3/4 = 2$), but the other current of spin 1 cannot
break spontaneously due
to Coleman theorem~\cite{coleman}, and indeed we find it again in
the $Z_4$ usual parafermion.
As the stress-energy tensor can be related to this current via $U(1)$
Sugawara construction, this implies also that conformal symmetry cannot be
broken along this ``flow'' and the central charge should not change, as indeed
is the case. All these strange features indicate that $K=4$ is somewhat a
``limiting'' case for $Z_K$ exotic superconformal algebras.

\vskip 0.3cm
{\bf $\SZ_5$ algebra -} In this case there are four parafermions of spin
$\D_1=\D_4=6/5$, $\D_2=\D_3=M_2-6/5$ respectively. The algebra has 3
parameters: $c$, $\lambda=C_{1,1}$ and $\mu=C_{1,2}$. The requirement that $R
\geq 0$ for all 4-point functions translates again into a set of inequalities
for $M_2$ that can be simultaneously satisfied only if $M_2=2,3$. In the first
case $M_2=2$ we have $\D_1=6/5$, $\D_2=4/5$ and we reobtain the usual $Z_5$
parafermionic model of Fateev and Zamolodchikov~\cite{fz}, merely with the
name of the two parafermions reversed. The other case is $\D_1=6/5$,
$\D_2=9/5$. Using the correlation function $\corr{1}{\uno}{1}{\uno}$,
with $R=3$, we
can fix $c=-6$ and $\lambda=\sqrt{4/5}$. The $\corr{1}{\uno}{2}{\due}$
correlation function, with $R=3$, reads
\eq
\corr{1}{\uno}{2}{\due} =
x^{-12/5}(1-x)^{-6/5} \left(1-\frac{6}{5}x-\frac{3}{5}
x^2+\frac{4}{5}x^3 \right)
\en
where we have fixed the coefficient of $x^2$ in the polynomial inserting $c=-6$
in eq.(\ref{c}), and that of $x^3$ by imposing {\em su}-duality, eq.(\ref{t}).
{\em st}-duality then requires that the sum of coefficients in the polynomial
$P(x)$ is equal to $|\mu|^2$. But this sum is zero, thus
implying $\mu=0$ in contradiction with the fusion rules. This argument rules
out the $M_2=3$ case.

$\SZ_5$ can then only coincide with the $Z_5$ parafermionic algebra with
$c=8/7$, $\lambda=\sqrt{8/5}$, $\mu=\sqrt{9/5}$. This is a very particular
case: it is easy to check that no other $Z_K$ parafermionic model contains
a current of spin $1+1/K$. See also appendix A for a more general result
on this point.

\vskip 0.3cm
{\bf $\SZ_{6,7}$ algebras -} For $K=6$ we have 5 parafermions of spin
$\D_1=\D_5=7/6$,
$\D_2=\D_4=M_2-4/3$, $\D_3=M_3-3/2$. The parameters in the algebra are $c$,
$\lambda=C_{1,1}$, $\mu=C_{1,2}$ and $\rho=C_{2,2}$. The requirement $R\geq 0$
for all 4-point functions restricts $M_2$ and $M_3$ to the following chioces:
\[
\begin{array}{lllll}
\mbox{A.} & M_2=M_3=2     & \D_1=7/6, & \D_2=2/3, & \D_3=1/2 \\
\mbox{B.} & M_2=2,~M_3=3  & \D_1=7/6, & \D_2=2/3, & \D_3=3/2 \\
\mbox{C.} & M_2=M_3=3     & \D_1=7/6, & \D_2=5/3, & \D_3=3/2
\end{array} \]
Case A is immediately ruled out because \{$1,\psi_2,\psi_4$\} form in this
case
a $Z_3$ parafermion subalgebra with $c=4/5$ while \{$1,\psi_3$\} form a $Z_2$
algebra with $c=1/2$, incompatible with the former.
Case B also is nonsense, as \{$1,\psi_2,\psi_4$\} still forms a
$Z_3$ subalgebra
with $c=4/5$, but we know that at $c=4/5$ there is no room for Virasoro
primary fields of dimensions 7/6 or 3/2.
We are left with case C, where the correlation function
$\corr{1}{\uno}{1}{\uno}$
(with $R=3$) fixes $c=-49/10$. Besides the fact that this automatically
excludes possibility to build up any unitary model, the $\SZ_6$ algebra of
case C is inconsistent even for this negative value of $c$. Indeed it can be
checked that the correlation functions $\corr{1}{\uno}{2}{\due}$ and
$\corr{1}{\uno}{3}{3}$ give two incompatible (and negative!) values of
$|\mu|^2$.

Thus there is no possibility to realize an associative $\SZ_6$ algebra.
The same
happens for $K=7$, where the only possibility $\D_1=8/7,\D_2=11/7, \D_3=9/7$
gives a negative value of $c$ and is ruled out by arguments similar to those
of $K=6$.
For higher values of $K$ the analysis is in principle still possible,
but the number of correlation functions to analyze increases and computations
can become cumbersome.
To get more general results we have to turn to a closer analysis
of some particular correlation function and restrict more the problem by
some new input.

\section{General unitary case - Proof of unconsistency for $K>5$}

If we are content to explore the possibility
to have {\em unitary} $Z_K$ exotic superconformal theories, then the
additional requirements $c>0$ and
$\D_i>0$ help to get a general answer. This can be given in form of a

{\bf Theorem -} {\em With the constraints $c>0$ and $\D_i>0$, $i=1,2,3,4$,
there is no associative $\SZ_K$ algebra for $K>5$}.

In the proof of this statement, we make use of three conformal dimensions
\eq
\D_1=1+\frac{1}{K} \virg \D_2=M_2-2+\frac{4}{K} \virg \D_3=M_3-3+\frac{9}{K}
\en
We already know from previous section that $\SZ_K$ algebras exist for $K\leq 5$
and are absent for $K=6,7$. Here we shall consider values of $K\geq 6$.
Reqirement $\D_i>0$ bounds $M_i$ to
\eq
M_2 \geq 2 \virg M_3 \geq \left\{ \begin{array}{ll}2 &\mbox{for $K\leq 8$}\\
                                                   3 &\mbox{for $K\geq 9$}
                                  \end{array} \right.
\en
Let us first consider the correlation function $\corr{1}{\uno}{1}{\uno}$, for
which $R=4\D_1-\D_2=6-M_2\geq 0$ reqires $M_2\leq 6$. The case $M_2=6$ leads
to $R=0$ and, as explained in section 3, is unconsistent, the same happens
for $M_2=5$, $R=1$ as $\alpha_{11}=M_2-4+\frac{2}{K}\not= 1$. The case $M_3$,
$R=3$ always gives, through eq.(\ref{c}) negative values of $c$, hence it is
discarded too. We are left with two possibilities:
\[
\begin{array}{lll}
\mbox{A.} & M_2=2, R=4 ~\Rightarrow & P(x)=1-2\frac{K+1}{K}x+P_2x^2
                                           -2\frac{K+1}{K}x^3+x^4 \\
\mbox{B.} & M_2=4, R=2 ~\Rightarrow & P(x)=1-\frac{2}{K}x+x^2 ~\Rightarrow
                                      c=\frac{2(K+1)}{K-2}
\end{array}
\]

{\em Case} A - To go on we have to resort to another correlation function,
namely to $\corr{1}{\uno}{2}{\due}$, that, for $M_2=2$, has $R=4-M_3$. The
reqirement $R > 0$ ($R=0$ again is unconsistent) then restrict $M_3\leq 3$.
Thus for $K\geq 9$, $M_3=3$, $R=1$ is the only possible value. In this case
$P(x)=1+\frac{4-K}{K}x$ and $|C_{11}|^2=P_1=\frac{4-K}{K}$ is negative for
all $K>4$. This rules out this case. There is still the possibility of $M_3=2$
for $K=6,7,8$. The cases $K=6,7$ are ruled out by the results of the previous
section. $K=8$ is the only case where we have also to consider $\D_4=M_4-2$.
$\D_4>0$ implies $M_4>2$ while the correlation function $\corr{2}{2}{2}{2}$
requires $M_4\leq 2$. Also this case is ruled out.

{\em Case} B - The function $\corr{1}{\uno}{1}{\uno}$ previously considered
is completely fixed in this $R=2$ case and yelds
\eq
c=\frac{2(K+1)}{K-2} \virg |C_{11}|^2=2+\frac{2}{K}
\label{B}
\en
In this case the correlation function $\corr{1}{\uno}{2}{\due}$ has $R=8-M_3$,
hence it must be $M_3<8$. Here it is convenient to resort to the
function $\corr{1}{1}{1}{\hat{3}}$, having $R=M_3-3M_2+6=M_3-6$, that
reqires $M_3\geq 6$. Hence $M_3=6,7$. Consider again the function
$\corr{1}{\uno}{2}{\due}$. For $M_3=7$, $R=1$ it is possible to compute
$|C_{11}|^2=1+\frac{4}{K}$, in contradiction with (\ref{B}). For $M_3=6$ the
value of $|C_{11}|^2$ in (\ref{B}) helps to fix the coefficient $P_2$ from
which $c$ can be evluated back. We obtain $c=\frac{4(K+1)(K+2)}{K^2-4K+8}$,
in contradiction with (\ref{B}).

All the possible cases are then ruled out by simply considering a set of
few correlation functions for some of the fields in the algebra. We believe
that this result, surely valid for unitary theories, is in fact absolutely
general: there are no $Z_K$ exotic (in the sense of $Q^K=P$) superconformal
algebras for $K>5$.

\section{Conclusions and implications}

The main result of this paper is the impossibility to construct $Z_K$ exotic
$N=1$ superconformal algebras for $K>5$. For $K=5$ the result is trivial,
for $K=4$ is $c=1$ theory, and finally we are ``seriously'' left only with
the already known cases $K=2,3$, i.e. ordinary $N=1$ superconformal algebra
and the spin 4/3 algebra. What is established is that for $K>5$ it is not
possible to realize at the conformal point, an algebra of currents such that
it can define a charge $Q$ satisfying $Q^K=P$.
This does not mean that ``more exotic''
supersymmetric algebras can not be constructed in two dimensions: for example,
Fateev~\cite{fat}
has recently shown that in some parafermionic models it is possible to
consider a charge $Q$ such that $Q^K=P_s$, where $P_s$ is
an appropriate higher spin local conserved charge. These models are in
connection with the parafermionic algebras introduced in the appendix A
of~\cite{fz}, and further studied, from the unitarity point of view,
in~\cite{gs}. In fact, $\SZ_3$ is also a particular case of these algebras.

As $Q^K=P$ can not be realized at criticality, it is presumably impossible
also in perturbations of CFT's. It then becomes problematic to identify
the Boltzmann weights of~\cite{bp} with the scattering matrices of a 2D QFT.
It can happen that such a QFT does not exist, but also that it exists and its
UV limit is quite tricky. Also, the spontneous symmetry breakdown and the
goldstino problem needs more understanding. An intriguing observation in this
connection is that the goldstino picture breaks down at $K=4$.
Palla~\cite{palla} has
studied perturbations of $Z_K$ parafermionic models that can convert some
of the parafermionic currents into conserved quantities. Now, this is
possible exactly starting from $K=4$. Many indications point to the fact
that the behaviour of parafermionic theories should be quite different for
$K<4$ and $K>4$, with $K=4$ as a limiting case. If this is related or not
to more fundamental issues like Galois theory is to be understood.

\vskip 0.5cm
{\bf Acknowledgements -} I am grateful to M.Bauer, D.Bernard, V.Pasquier
and J.B.Zuber for many useful discussions.
I thank the {\em Service de Physique
Th{\'e}orique} of C.E.A. - Saclay for the kind ospitality. The Theory Group of
I.N.F.N. - Bologna
and the Director of Sez. di Bologna of I.N.F.N. are acknowledged for the
financial support allowing me to spend this year in Saclay.

\section*{Appendix}

It is interesting to ask if there are $\SZ_P$ algebras hidden in the usual
$Z_K$ parafermionic models, even for $P\not= K$, generalizing the curious
phenomenon observed for $Z_5$. We shall show here that no such case is
possible for $P>5$, in agreement with the results of the main part of the
paper, and that the only other cases can always be traced back to the well
known $\SZ_2$ or $\SZ_3$ algebras.

First of all let us prove that for $P>5$ there is no parafermion of spin
$\frac{P+1}{P}$ contained in any $Z_K$ parafermionic algebra, for all $K$.
To do that, we have to consider the equation
\eq
\frac{P+1}{P}=\frac{k(K-k)}{K}
\label{P}
\en
where the expression on the r.h.s., for $k=1,...,K-1$ gives the most general
spin for a parafermion in $Z_K$. The case $k=1$ is clearly impossible, for
all $K$ and all $P$ as it equates a l.h.s. greater than 1 with a r.h.s. less
than 1. So consider $k\geq 2$. Solving (\ref{P}) for $K$ one gets
\eq
K=k+1+\frac{k+P+1}{Pk-P-1}
\label{K}
\en
As $K$ must be an integer, we have to reqire $k+P+1\geq Pk-P-1$, i.e.
\eq
k\leq 2\frac{P+1}{P-1}
\label{diseg}
\en
For $P>5$ this implies $k\leq 2$, hence the only possibility is $k=2$.
Substituting in (\ref{K}) we get
\eq
K=4+\frac{4}{P-1}
\en
which can never be integer if $P>5$. This proves that no $\SZ_P$ is contained
in a usual parafermionic algebra for $P>5$.

For $P\leq 5$ use of (\ref{K}) and (\ref{diseg}) allows to list all the
possible occurrencies.
\begin{itemize}
\item
For $P=5$ we have $K=5$, $k=2,3$. This is the result noticed in the paper
that $\SZ_5$ coincides with the $Z_5$ parafermion.
\item
For $P=4$ no solution appears.
\item
For $P=3$ we have $K=6$, $k=2,4$, thus showing that a realization of $\SZ_3$
is contained in the $Z_6$ parafermionic algebra. This is not surprising as
the $Z_6$ model is known to belong to the unitary minimal series of spin 4/3
algebra, namely for $m=4$.
\item
For $P=2$ there are two solutions: $K=6$, $k=3$ says that the $Z_6$ model is
also supersymmetric, (it belongs indeed also to the superconformal minimal
series for $m=6$) while $K=8$, $k=2,6$ shows {\em two} fields of spin 3/2
for the $Z_8$ model.
\item
For $P=1$ the solution $K=8$, $k=4$ completes the result for $P=2$: the two
spin 3/2 curents are associated to a current of spin 2, thus \{$T,\psi_2,
\psi_4,\psi_6$ form in this case two copies of the $N=1$ superconformal
algebra: the $Z_8$ model is doubly $N=1$ supersymmetric. Another solution
appears for $P=1$ when $K=9$, $k=3,6$.
\end{itemize}
This exhausts all possible realizations of $\SZ_P$ algebras in $Z_K$
parafermionic models.

\end{document}